\documentclass{emulateapj}
\usepackage{bm}
\usepackage{chngpage}
\usepackage{graphicx}
\usepackage[caption=false]{subfig}
\usepackage{mathrsfs}

\newcommand{\degree}{\ensuremath{^\circ}}
\def\apjl{ApJL}

\shorttitle{Relativistic MHD Simulation of Jets in GRBs}
\shortauthors{Geng et al.}

\begin{document}

\title{Propagation of a Short GRB Jet in the Ejecta: Jet Launching Delay Time, Jet Structure, and GW170817/GRB 170817A}

\author{Jin-Jun Geng\altaffilmark{1,2,3}, Bing Zhang\altaffilmark{4}, Anders K\"olligan\altaffilmark{5}, Rolf Kuiper\altaffilmark{5},
       Yong-Feng Huang\altaffilmark{1,2}}

\altaffiltext{1}{School of Astronomy and Space Science, Nanjing University, Nanjing 210023, People's Republic of China;
                 gengjinjun@nju.edu.cn, hyf@nju.edu.cn}
\altaffiltext{2}{Key Laboratory of Modern Astronomy and Astrophysics (Nanjing University), Ministry of Education, Nanjing 210023, People's Republic of China}
\altaffiltext{3}{Department of Physics, Nanjing University, Nanjing 210093, People's Republic of China}
\altaffiltext{4}{Department of Physics and Astronomy, University of Nevada, Las Vegas, NV 89154, USA; zhang@physics.unlv.edu}
\altaffiltext{5}{Institute of Astronomy and Astrophysics, University of T\"ubingen, Auf der Morgenstelle 10, D-72076 T\"ubingen}

\begin{abstract}
We perform a series of relativistic magnetohydrodynamics simulations to investigate how a hot magnetic jet propagates within the dynamical ejecta of a binary neutron star merger, focusing on how the jet structure depends on the delay time of jet launching with respect to the merger time, $\Delta t_{\rm jet}$. We find that regardless of the jet launching delay time, a structured jet with an angle-dependent luminosity and Lorentz factor is always formed after the jet breaks out of the ejecta. On the other hand, the jet launching delay time has an impact on the jet structure. If the jet launching delay time is relatively long, e.g., $\ge 0.5$ s, the line-of-sight material has a dominant contribution from the cocoon. 
On the other hand, for a relatively short jet launching delay time, the jet penetrates through the ejecta early on and develops an angular structure afterward. The line-of-sight ejecta is dominated by the structured jet itself. 
We discuss the case of GW170817/GRB 170817A within the framework of both long and short jet launching delay time.
In the future, more observations of GW/GRB associations can help to differentiate between these two scenarios. 
\end{abstract}


\section{Introduction}

Gamma-ray bursts (GRBs) are believed to originate from a relativistic jet launched by a compact central engine, either a black hole (BH) or a rapidly rotating, highly magnetized neutron star (NS). After being launched, the jet propagates through and breaks out of the surrounding material before emitting $\gamma$-ray photons at large radii. The jet propagation physics has been investigated by many authors using numerical simulations, both within the context of long GRBs for the envelope of a massive progenitor star \citep[e.g.,][]{MacFadyen99,Aloy00,Zhang03,Morsony07,Mizuta09,Mizuta13,Lopez16,Geng16}, and within the context of short GRBs for the dynamical ejecta of a binary NS merger \citep{Nagakura14,Murguia14}. Relevant analytical studies have also been arried out \citep{Bromberg11,Bromberg14}. These studies show that the interaction between the jet and the surrounding material produces a hot cocoon surrounding the jet, which in turn helps to collimate the jet.

The coincident detection of a gravitational wave (GW) event GW170817 \citep{Abbott17} and a short gamma-ray burst (GRB) 170817A \citep{Goldstein17,ZhangBB18} confirms the hypothesis that binary NS mergers are the progenitors of at least some short GRBs. In the literature, a uniform conical jet with a sharp edge (also called a top-hat jet) was usually engaged to interpret GRB prompt emission and afterglows. However, such a simple model fails to explain the prompt emission data of GRB 170817A \citep[e.g.,][]{Granot17,Gottlieb18,Meng18}. The brightening of the late-time X-ray/optical/radio afterglow hints the existence of a significant energy injection, which was interpreted as either lateral injection from a structured jet 
\citep{Kathirgamaraju18,Lazzati18,Lamb18a,Lamb18b,Lyman18,Xie18,Ghirlanda19,Troja18,Piro19,Xiao17} or radial injection from a stratified cocoon \citep{Kasliwal17,Mooley18b,Nakar17,Margutti18}. The detection of ``superluminal'' motion in the radio afterglow \citep{Mooley18a,Ghirlanda19} ruled out the later scenario and reinforced the structured jet picture. However, it is unclear whether the line-of-sight material, which moves with a mild Lorentz factor, comes from the cocoon surrounding the central jet or from the wing of the structured jet itself. 

Observationally, GRB 170817A is delayed with respect to GW170817 by $\Delta t \sim 1.7$ s. The origin of this delay has not been clarified~\citep{Zhang19,Gill19}. Some authors \citep[e.g.,][]{Gottlieb18,Bromberg18} attributed a significant portion of this delay to the delay of jet launching, $\Delta t_{\rm jet}$. Under such an assumption, the dynamical ejecta already propagates to a distance of $\sim v_{\rm ej} \Delta t_{\rm jet} \sim 6\times 10^9 \ {\rm cm} \ (v_{\rm ej}/0.2 c) (\Delta t_{\rm jet}/1 {\rm s})$ when the jet is launched, where $v_{\rm ej}$ is the average velocity of the dynamical ejecta. The interaction between the jet and this extended ejecta is significant, making a significant cocoon component. The prompt $\gamma$-ray emission may be explained as the photospheric emission of the cocoon as the jet breaks out of the ejecta \citep{Lazzati17,Nakar18}. On the other hand, \cite{ZhangBB18} pointed out that observationally the delay time $\Delta t \sim 1.7$ s is comparable to the duration of the burst $T_{90} \sim 2$ s, which is consistent with the scenario that both time scales are defined by the same physical quantity $\sim R_{\rm GRB}/\Gamma^2 c$. This suggests that the delay time scale is mostly defined by the time when the jet propagates to the dissipation site at $R_{\rm GRB}$, which is much greater than the photosphere radius for typical parameters. Within this framework, $\Delta t_{\rm jet}$ is negligibly small. It is thus interesting to investigate the interaction between the jet and the dynamical ejecta, considering the effect of $\Delta t_{\rm jet}$ in general. This is one of the main goals of this paper. 

Within the context of short GRBs and in particular the GW170817/GRB 170817A association, relativistic hydrodynamical simulations have been performed to explore how the jet power and Lorentz factor vary as a function of the polar angle \citep[e.g.,][]{Lazzati18,Xie18,Wu18}, and the propagation of a Poynting-flux-dominated jet was presented using relativistic magnetohydrodynamic (RMHD) simulations by \cite{Bromberg18}.
Lately, \cite{Kathirgamaraj19} and \cite{Fernandez19} investigated the jet structure by considering the jet launching mechanism from the central engine. The density contributions from the post-merger and dynamical ejecta are not considered in the setups. In all these previous work, 
the role of $\Delta t_{\rm jet}$ was not investigated in detail.

In this paper, we perform a set of axisymmetric 2.5-dimensional RMHD simulations to study the propagation of a hot magnetic short GRB jet through the ejecta and after its breakout.  We pay special attention on the role of $\Delta t_{\rm jet}$ in defining the angular structure of the luminosity and Lorentz factor. The simulation setup is presented in Section 2. The simulation results are presented in Section 3, and their application to GW170817/GRB 170817A is discussed in Section 4. Our findings are summarized in Section 5.

\section{Simulation Setup}

Numerical simulations of binary NS mergers indicate that about $10^{-4}-10^{-2}$ solar masses ($M_{\odot}$) of NS material, called dynamical ejecta, are ejected during the coalescence \citep{Hotokezaka13,Shibata17}. In our simulation of jet propagation, the setups include a proper description of both the dynamical ejecta and the jet itself, which is given in details below.

\subsection{The Dynamical Ejecta}

In previous simulations of jet propagation, the ejecta is usually set up according to an initial condition, i.e., a certain density/velocity profile of the ejecta is already set before the jet launching in the simulation domain. On the other hand, the simulations of binary NS mergers show that essentially all the ejecta materials are ejected within $\sim$ 15 ms, regardless of the equation of state of the NS \citep{Hotokezaka13}. 
Since we are investigating how $\Delta t_{\rm jet}$ affects jet propagation, we adopt a more realistic treatment by setting the inner boundary condition of the ejecta that lasts for 15 ms, and let the ejecta propagate for $\Delta t_{\rm jet}$ before the jet is launched.
The average velocity of the ejecta, $v_{\rm ej}$, is assumed to have a typical value $\sim 0.2~c$. The density of the ejecta is set to have an angular profile, i.e., denser near the equator and more dilute near the axis. The initial density profile of the ejecta is set to (see \citealt{Kasliwal17,Xie18}),
\begin{equation}
\rho (t, \theta) = \left\{
\begin{array}{ll}
\rho_{\rm ej} (\frac{1}{4} + \sin^3 \theta),~~~~~~& t < 5~\mathrm{ms}, \\
\rho_{\rm ej} \left(\frac{t}{5~\mathrm{ms}}\right)^{-2} (\frac{1}{4} + \sin^3 \theta),~~~~~~& 5~\mathrm{ms} < t < 15~\mathrm{ms},
\end{array}\right.
\end{equation}
where $\theta$ is the polar angle measured from the jet axis, $\rho_{\rm ej}$ is solved from the ejecta mass ($M_{\rm ej}$) by 
$\int_{0}^{15~\mathrm{ms}} \rho(t, \theta) v_{\rm ej} r_{\rm in}^2 dt d \Omega = M_{\rm ej}$.
Similar to other works, the inner boundary of the simulation domain $r_{\rm in}$ is set to be $5 \times 10^7$~cm,
and $M_{\rm ej} = 10^{-2} M_{\odot}$ is adopted.

\subsection{The Jet}

Since we mainly focus on the jet propagation, in our simulations a relativistic jet is produced via a set of boundary conditions without including the detailed jet-launching mechanism. We consider a stable, hot, magnetic jet, of which the transversal equilibrium between the total pressure gradient, the centrifugal force, and the magnetic tension is already established. The jet material at the inlet is characterized by eight angular functions, namely density ($\rho_j$) and pressure ($p_j$) in the fluid frame, velocities in three directions ($v^r_j$, $v^{\theta}_j$, $v^{\phi}_j$) and magnetic field in the laboratory frame ($B^r$, $B^{\theta}$, $B^{\phi}$). These functions are determined as follows.

From an observational point of view, a jet is described by its luminosity ($L_j$) and the terminal Lorentz factor $\Gamma_\infty$. 
The specific gas enthalpy of the jet material is given by
\begin{equation}
h = 1 + \frac{1}{\rho_j c^2} \left( \langle e_j \rangle+ \langle p_j \rangle \right)
= 1 + \frac{4 \langle p_j \rangle}{\rho_j c^2},
\label{eq:h}
\end{equation}
where $e_j$ is the internal energy density and we have taken the adiabatic index as 4/3 ($e_j = 3 p_j$).
The brackets in Equation (\ref{eq:h}) denote the average values across the half-opening angle of the jet $\theta_j$.  
Defining two magnetization parameters as\footnote{Notice that here we adopt a more generalized definition of $\sigma$ by taking the enthalpy rather than the rest mass density in the denominator. Note that the total co-moving magnetic energy density can be written as 
$b^2 = b^{\mu} b_{\mu} = \frac{\bm{B}^2}{\Gamma^2} + (\bm{v} \cdot \bm{B})^2
\approx \frac{(B^{r})^2+(B^{\phi})^2}{\Gamma_r^2} + (B^r \cdot v_j^r)^2 = (B^{r})^2 + \frac{(B^{\phi})^2}{\Gamma_r^2} = (B^{r})^2 + (b^{\phi})^2$
when $B^{\theta} = v_j^{\theta} = 0$ and $v_j^{\phi} < v_j^{r}$ ($\Gamma \approx \Gamma_r = 1/\sqrt{1-(v_j^r)^2}$).}
\begin{equation}
\sigma_{\phi} \simeq \frac{\langle (b^{\phi})^2 \rangle}{2 \langle p_j \rangle},\sigma_r \simeq \frac{(B^r)^2}{2 \langle p_j \rangle},
\label{eq:sigma}
\end{equation}
where $(b^{\phi})^2$ is the energy density of the azimuthal magnetic field in the fluid frame, one can then define the specific enthalpy including the contribution of the magnetic field as
\begin{equation}
h^* = 1 + \frac{1}{\rho_j c^2} \left( \langle e_j \rangle+ \langle p_j \rangle \right) + \frac{b^2}{\rho_j c^2} = h + \frac{1}{2} (h-1) (\sigma_r + \sigma_{\phi}).
\end{equation}
When the internal energy and the magnetic energy in the jet are fully converted to the kinetic energy,
the jet material would reach a terminal Lorentz factor $\Gamma_\infty$, which is calculated as
$\Gamma_\infty \sim \Gamma_r \times h^*$ ($\Gamma_r$ is the initial Lorentz factor of the radial direction). 
On the other hand, the energy density in the lab frame can be expressed as
$e_{\rm lab} = \Gamma_r^2 h^* \rho_j c^2$,
which is related to $L_j$ by
\begin{equation}
L_j = 4 \pi r_{\rm in}^2 v^r_j e_{\rm lab}.
\end{equation}
For simplicity, we use a top-hat profile for all three parameters $\rho_j$, $v^r_j$ and $B^r$, 
which mean that they are constant within $\theta_j$.
Thus, $\rho_j$ could be obtained from Equation (5) under specific values of $L_j$, $\Gamma_\infty$, and $v^r_j$.

For the azimuthal magnetic field we adopt a profile as (also see \citealt{Marti15})
\begin{equation}
B^{\phi} (\theta) = \left\{
\begin{array}{ll}
\frac{2 B_{j,m}^{\phi} (\theta/\theta_m)}{1+(\theta/\theta_m)^2},~~~~~~& 0 < \theta < \theta_j, \\
0,~~~~~~& \theta > \theta_j,
\end{array}\right.
\label{eq:Bphi}
\end{equation}
where the azimuthal magnetic field increases linearly for $\theta \ll \theta_m$,  
reaching a maximum ($B_{j,m}^{\phi}$) at $\theta_m$ and decreasing as $1/\theta_j$ for $\theta > \theta_m$. 
A moderate value for $\theta_m / \theta_j$, 0.4, is adopted as a typical magnetic profile.
Since the central engine of a short GRB is believed to be rapidly rotating, the jet is assumed to be in rigid rotation, i.e., 
\begin{equation}
v^{\phi}_j (\theta) = v_{j,m}^{\phi} (\theta/\theta_j),~0 < \theta < \theta_j.
\label{eq:vphi}
\end{equation}
In the following simulations, $v_{j,m}^{\phi}$ is set to be $0.4 c$, 
which is equivalent of having a central engine with a rotating period of $\sim 5$ ms at the inner boundary of the jet;
$v^{\theta}_j$ and $B^{\theta}$ are set to be zero since they are usually much smaller than 
other components of an RMHD jet at the inlet. 
Combining Equations (\ref{eq:sigma}) and (\ref{eq:Bphi}-\ref{eq:vphi}) we then solve the transversal equilibrium equation (see \citealt{Marti15})
\begin{eqnarray}
\frac{d p_j}{d \theta} & = & -\frac{(B^{\phi})^2}{\theta (\Gamma^r)^2} - \left(\frac{B^{\phi}}{(\Gamma^r)^2} +  B^r v_j^r v^{\phi}_j \right) \frac{d B^{\phi}}{d \theta} \nonumber \\
& + & \left( B^r v^{\phi}_j - v_j^r B^{\phi}\right) B^r \frac{d v_j^{\phi}}{d \theta} + \frac{h \rho_j c^2 \Gamma^2 (v^{\phi}_j)^2}{\theta} \nonumber \\
& + & \frac{B^r v^{\phi}_j}{\theta} \left( B^r v^{\phi}_j -  2 B^{\phi} v^r_j \right)
\end{eqnarray}
to obtain $p_j$ with specific $B^r$ and $B_{j,m}^{\phi}$.
In practice, $B^r$ and $B_{j,m}^{\phi}$ are fixed to meet the conditions of $\sigma_r$ and $\sigma_{\phi}$ through several trials.

The magnetization of the GRB jet is still under debate. 
The initial magnetization may be high ($\ge 100$) in the vicinity of the central engine.
On the other hand, dissipation processes may convert a significant fraction of the magnetic energy into the internal energy 
soon after launch \citep{Bromberg16}. In this paper, we choose an equipartition case,
i.e., $\sigma_r \sim \sigma_{\phi} \sim 1$ in the simulations. 

After the NS merger, the jet launching may be delayed by a duration of $\Delta t_{\rm jet}$
in comparison with the start time of ejecta. 
The value of $\Delta t_{\rm jet}$ depends on the type of the central engine and the jet launching mechanism of GRBs.
In general, $\Delta t_{\rm jet}$ should consist of the timescale to establish either an accretion disk
or strong magnetic fields, and the timescale to launch a relativistic jet when mass loading is low enough.
It may range from 10 ms to several seconds (see \citealt{Zhang19} for details).
In our simulations, we investigate
four representative values of $\Delta t_{\rm jet}$, i.e., 0.01 s, 0.1 s, 0.5 s, 1.0 s, respectively. 
The jet scenarios are named in the form of ``M$i$'', $i = \Delta t_{\rm jet}$.
We have chosen the same values of $\theta_j$, $L_j$, $\Gamma_{\infty}$ and $v_j^r$ for these four cases 
to isolate the effect of $\Delta t_{\rm jet}$. 
All the initial parameters of the four scenarios are listed in Table 1.
The jet angle $\theta_j$ is taken to be $10\degree$, which is roughly a median jet opening angle for short GRBs~\citep{Fong15,Beniamini19}.
A typical value of $L_j = 5 \times 10^{51}$ erg s$^{-1}$~\citep{Guetta05} and $\Gamma_{\infty} \sim 300$ are adopted.
In Table 1, one could see $B^r \sim B_{j,m}^{\phi} \sim 10^{12}$~G at $r_{\rm in}$.
This corresponds to a magnetic strength of $10^{14}$ G at $10^{6}$~cm (the surface of a central object),
and is consistent with that invoked in previous jet launching theories (e.g.,~\citealt{Blandford77}).

\subsection{Method}

We run axisymmetric and midplane symmetric 2.5-dimension RMHD simulations using the PLUTO code (version 4.2, see \citealt{Mignone07} for a full description). Spherical coordinates $(r,\theta)$ are employed and axisymmetry is assumed for all the simulations.
The computational domain covers a region of $r_{\rm in} \le r \le 6 \times 10^{10}~\mathrm{cm}$ and $0^{\circ} \le \theta \le 90^{\circ}$.
The radial grid consists of 2312 points and is logarithmically distributed, while the angular grid is uniform with 512 points,
making the cell aspect ratio to be $\sim 1$.
With this setup, the jet is resolved by roughly 60 cells across $\theta_j$, comparable to previous 2D studies.
A Riemann solver, called HLLD solver (see \citealt{Miyoshi05}), a linear-type spatial reconstruction, 
and a second-order Runge-Kutta time integration were chosen in the simulations.
As a result, we achieve the second-order accuracy in both space and time. 

\section{Simulation Results}

\begin{figure}
\centering
   \subfloat{\includegraphics[width=0.4\linewidth]{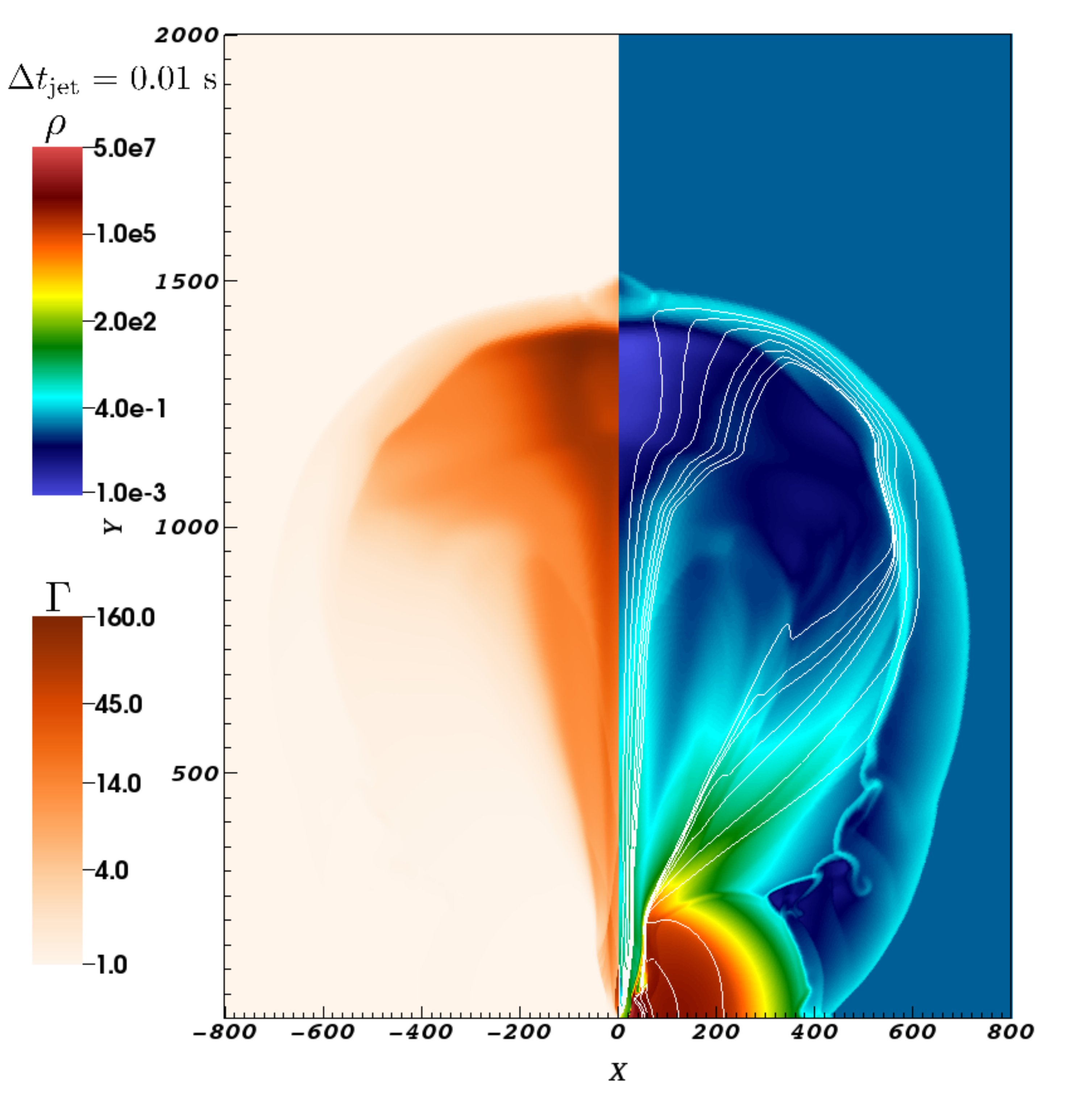}} 
   \subfloat{\includegraphics[width=0.4\linewidth]{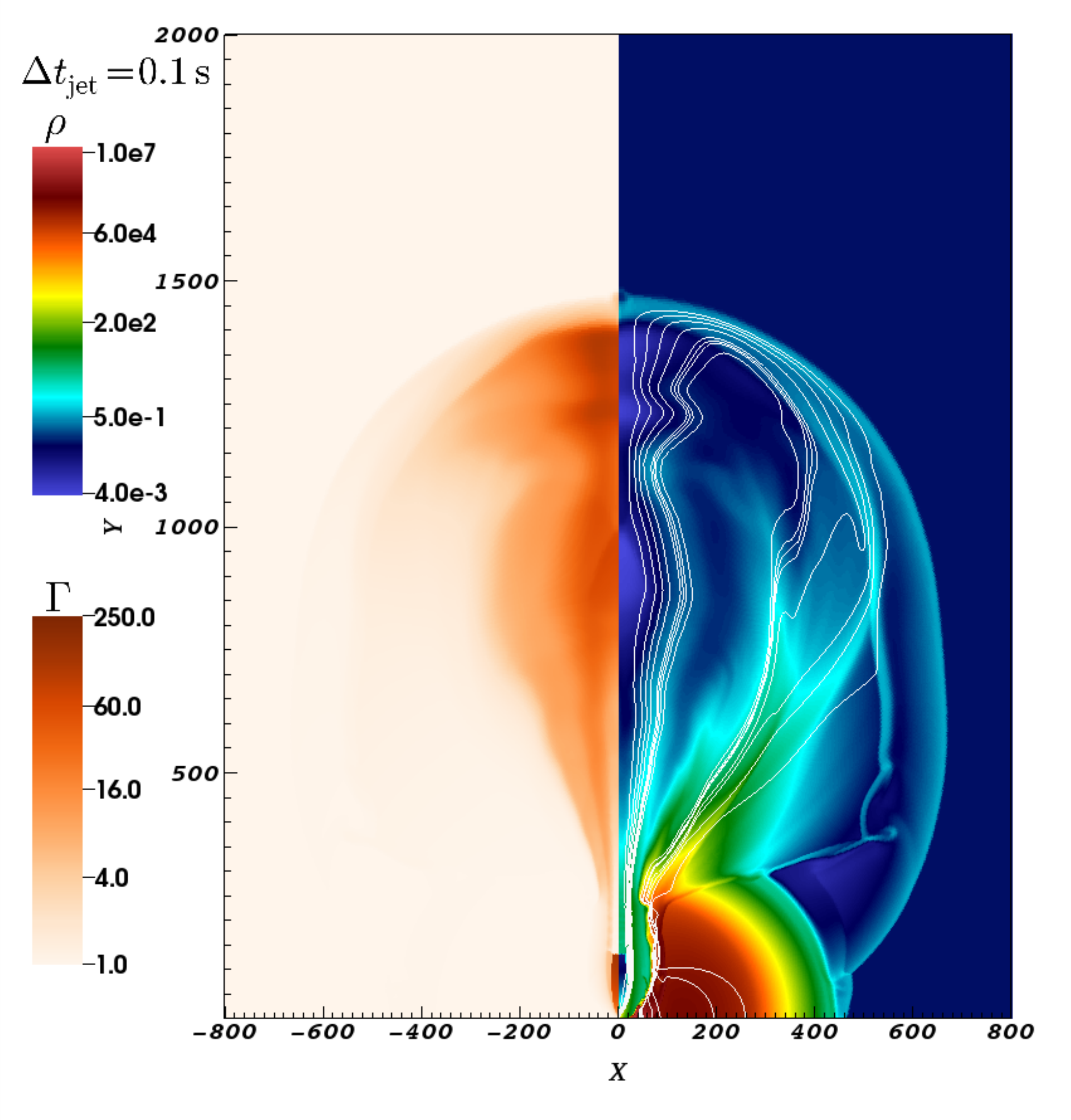}} \\
   \subfloat{\includegraphics[width=0.4\linewidth]{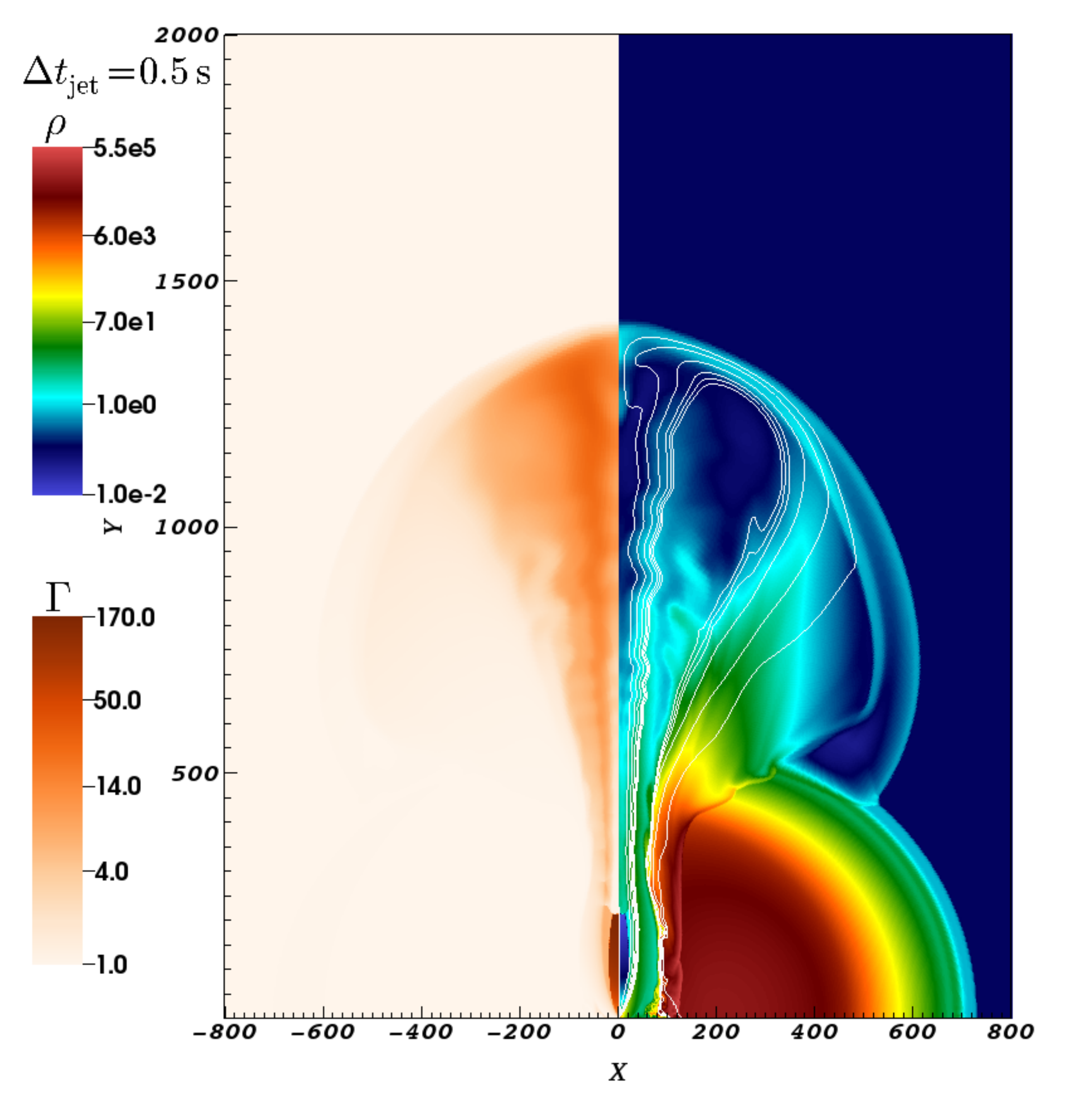}} 
   \subfloat{\includegraphics[width=0.4\linewidth]{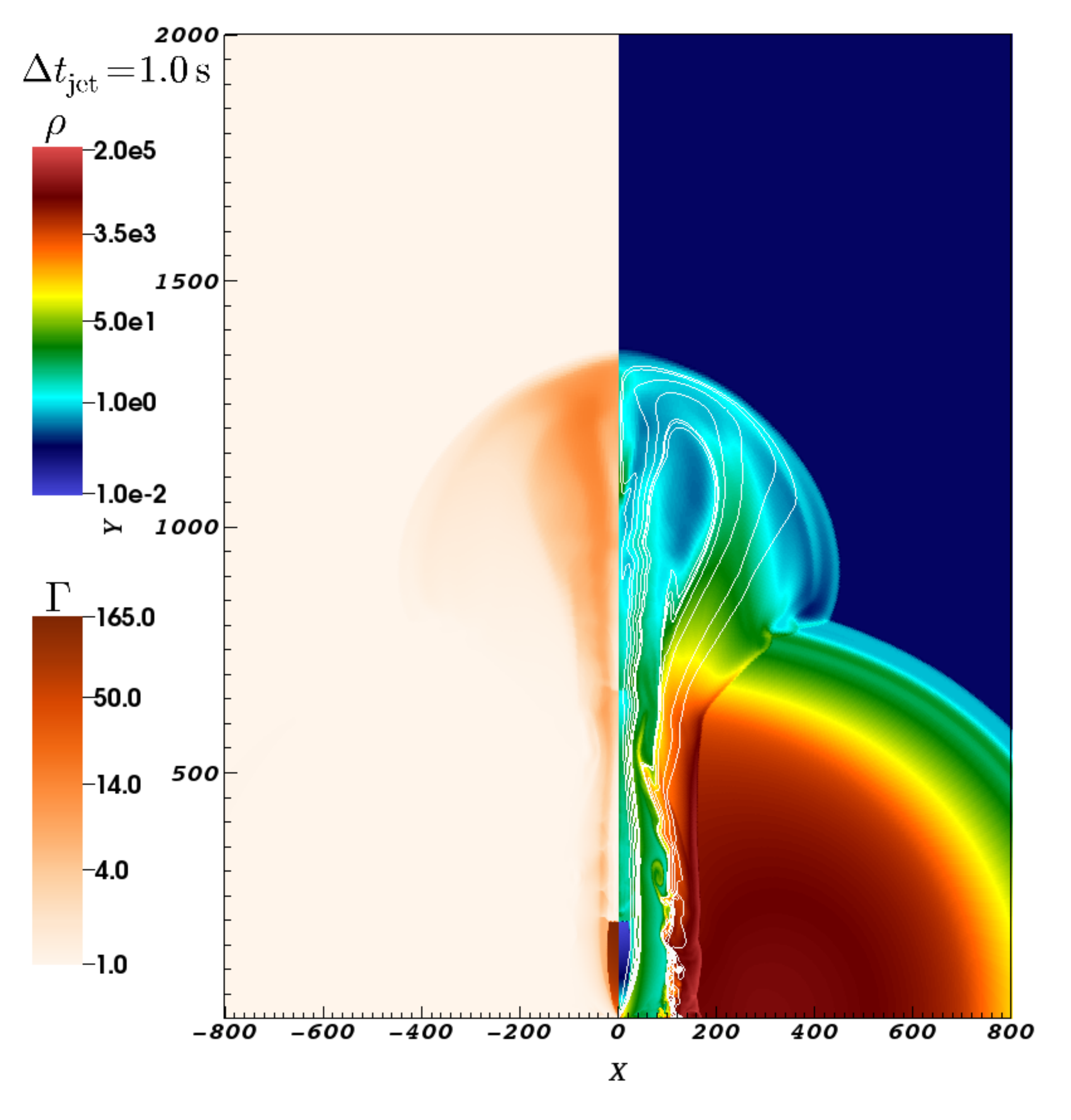}}
\caption{Lorentz factor map and density distribution (in units of g~cm$^{-3}$)
of four scenarios (with different $\Delta t_{\rm jet}$) when the jet material has been ejected for a period of 0.5 s.
Blue lines depict the magnetic field lines on the X-Y plane.
The unit scale for X and Y axis is $10^7$~cm.}
\label{Fig:rho}
\end{figure}

Before the jet launching, the dynamical ejecta is injected first as shown in Section 2.1.
After $\Delta t_{\rm jet}$, the jet material is injected as described in Section 2.2.
We have simulated four jets with the delay time of $\Delta t_{\rm jet} =$ 0.01 s, 0.1 s, 0.5 s, and 1.0 s respectively. 
The total duration of the jet is 1.0 s for all cases. 
In Figure \ref{Fig:rho}, we show the distributions of density and $\Gamma$ 
at the time when the jet has been launched for 0.5 s, i.e., half of the total duration.
Similar to pure hydrodynamical simulations, a cocoon emerges to generate the pressure needed to 
counterbalance the pressure from the surrounding ejecta. 
It is seen that the interaction between the jet and the cocoon is weaker
for a smaller $\Delta t_{\rm jet}$ since the jet funnel is formed quickly,
which leads to a higher $\Gamma$ for materials beyond the $\theta_j$.
After the breakout, the Lorentz factor of the jet core accelerates linearly with $r$,
while the Lorentz factor beyond the jet core becomes angle-dependent,
together with the expansion of the cocoon material.

\begin{figure}
   \centering
   \includegraphics[scale=0.4]{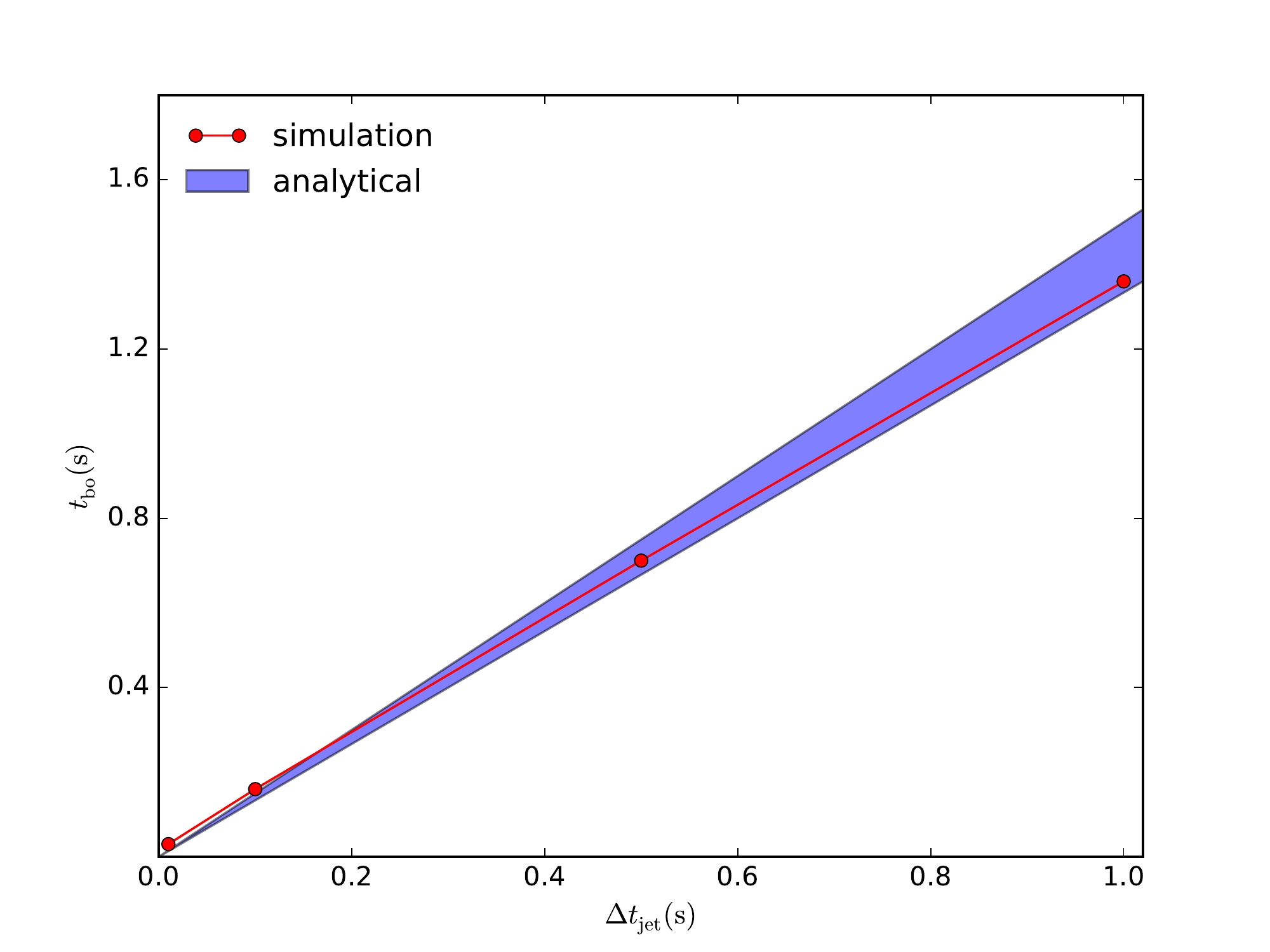}
   \caption{The relation between the breakout time of the jet and $\Delta t_{\rm jet}$
   for our simulation results. The shadow area shows the analytic results when the jet 
   head velocity is in the range of 0.3-0.4 c.}
   \label{fig:tbo}
\end{figure}

The breakout time of the hot magnetic jet, $t_{\rm bo}$ for each simulation is
presented in Figure \ref{fig:tbo}.
Assigning the average velocity of the jet head before breakout as $v_{\rm hj}$,
the jet breakout time since the merger can be estimated as
\begin{equation}
t_{\rm bo} = \Delta t_{\rm jet} + \frac{v_{\rm ej} \Delta t_{\rm jet}}{v_{\rm jh}-v_{\rm ej}}.
\label{eq:tbo}
\end{equation}
As shown in Figure \ref{fig:tbo}, $t_{\rm bo}$ obtained from the simulation
results is well consistent with that dervived from Equation \ref{eq:tbo}
when $v_{\rm jh}$ is within the range of 0.3 c to 0.4 c.

From these simulations, we investigate the relationship between $\Delta t_{\rm jet}$ 
and the structure of the jet.
The jet is quenched artificially at the inner boundary after being launched for 1 s.
At the time when the last injected jet material has escaped from the outer edge of the ejecta,
we can calculate the equivalent Lorentz factor averaged along the radial direction
as an estimate for the terminal Lorentz factor of the outflow, i.e.,
\begin{equation}
\bar{\Gamma} (\theta) = \left( \frac{\int_{r_{\rm out}} \Gamma^2(\rho c^2+4 p+b^2)~dV}{\int \rho c^2~dV} \right)^{1/2},
\end{equation}
where $r_{\rm out}$ is the radius of the outer edge of the ejecta.
One could compare the equivalent Lorentz factor for each case in a straightforward view
rather than a 2D view in Figure 1.
Similarly, one could derive the outflow energy per solid angle by integrating energy along the radial direction
\begin{equation}
\frac{d E}{d \Omega} = \frac{\int_{r_{\rm out}} \Gamma^2(\rho c^2+4 p+b^2)~dV}{2 \pi \sin \theta d \theta}.
\end{equation}

\begin{figure}
   \centering
   \includegraphics[scale=0.4]{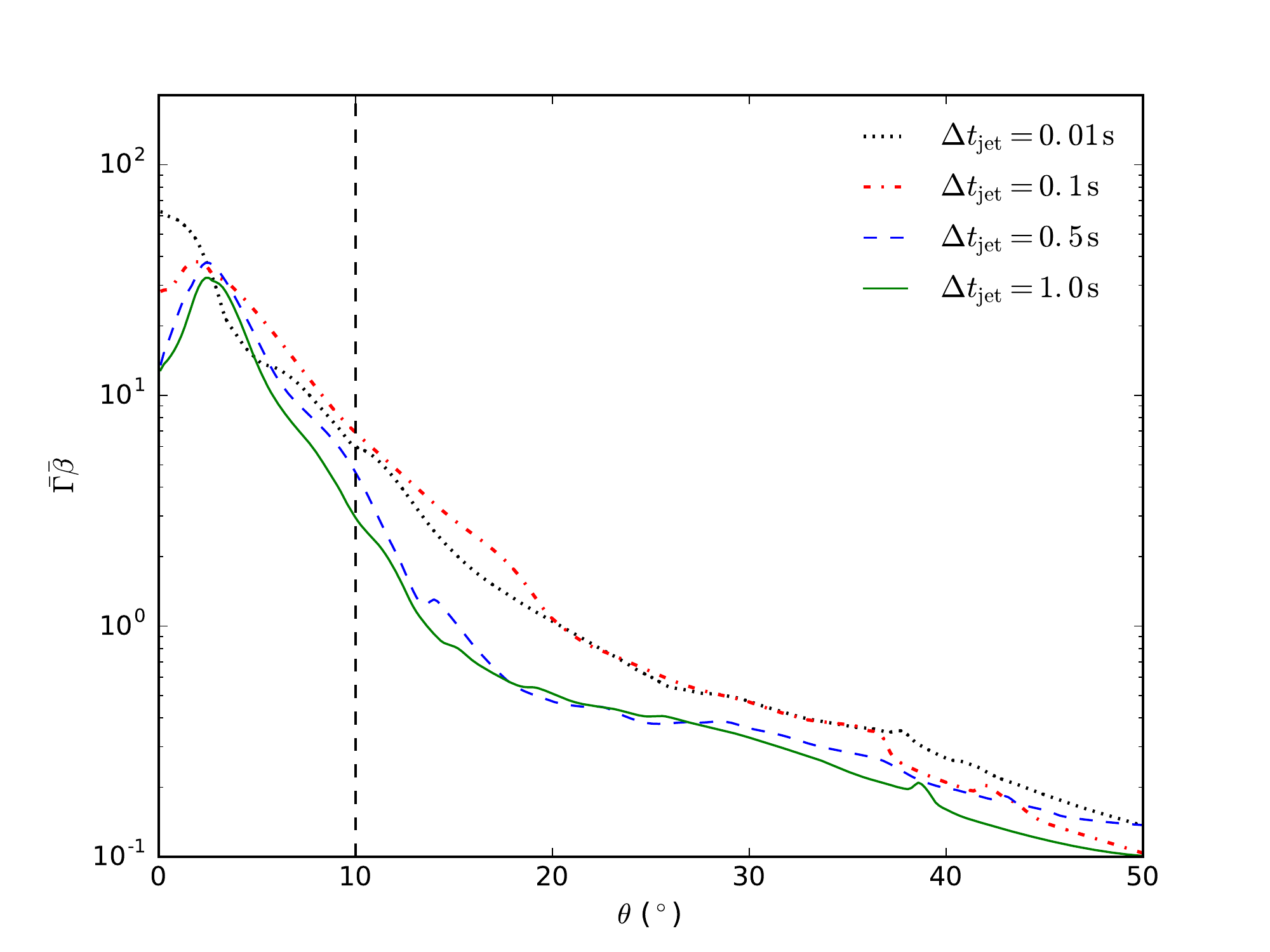}
   \caption{The angular distribution of $\bar{\Gamma} \bar{\beta}$ of the outflow for the four cases
   after the jet escapes from the outer edge of the ejecta.
   The position of initial $\theta_j$ is marked by the vertical dashed line.}
   \label{fig:Gamma}
\end{figure}

\begin{figure}
   \centering
   \includegraphics[scale=0.4]{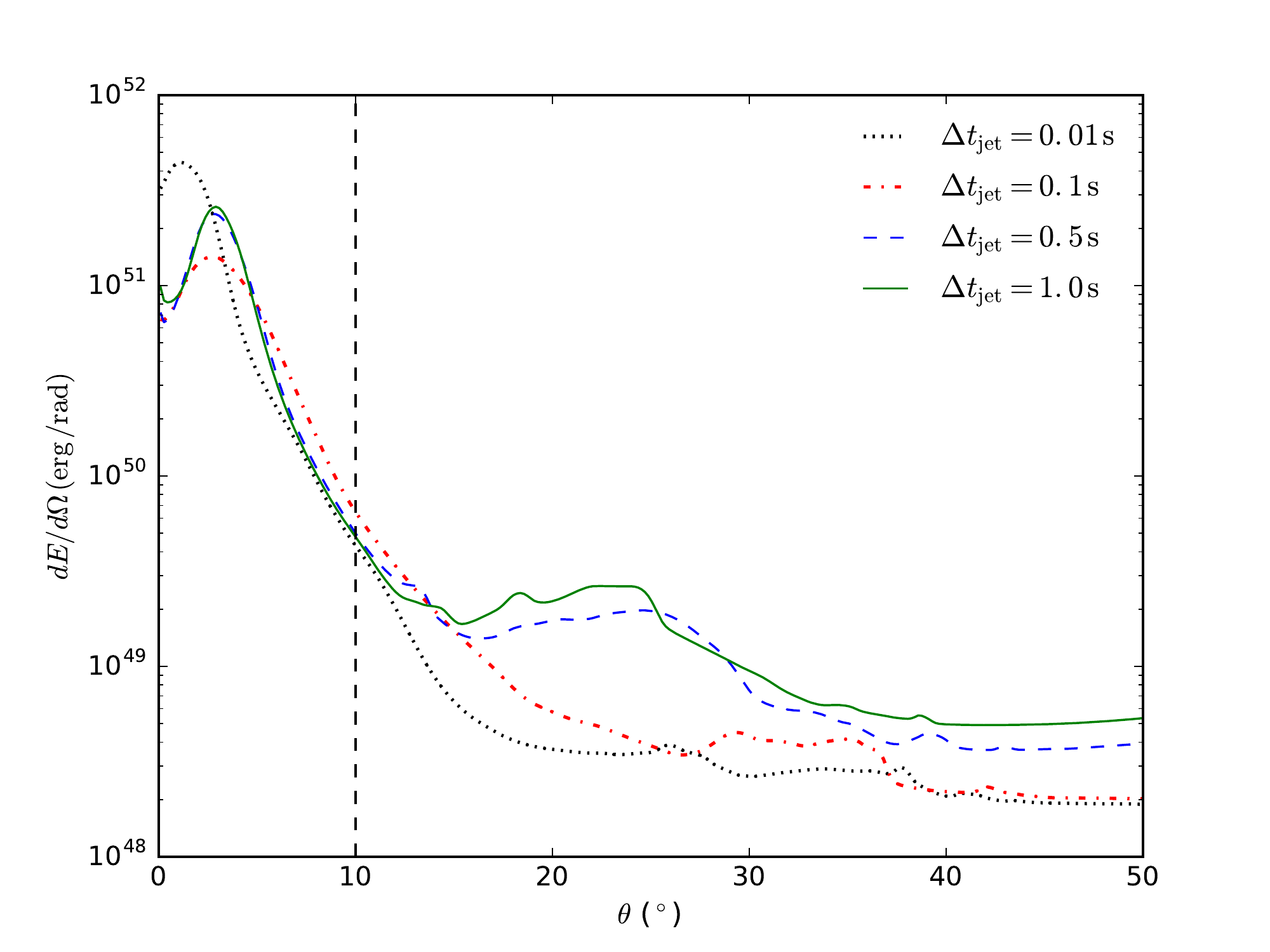}
   \caption{The angular distribution of the total energy per solid angle of the outflow for the four cases.}
   \label{fig:dE}
\end{figure}

The jet $\bar{\Gamma} \bar{\beta}$ and energy angular structure for the four jets are presented in
Figure \ref{fig:Gamma} and Figure \ref{fig:dE}, respectively.
One can draw the following interesting conclusions. 
First, regardless of $\Delta t_{\rm jet}$, an angular structure is always formed for both $\bar{\Gamma} \bar{\beta}$
and $dE / d\Omega$, and at the viewing angle $\theta_v$, which is several times of the jet opening angle $\theta_j$,
there is always mildly relativistic ejecta moving along the line of sight. The case of GW170817/GRB 170817A
is therefore naturally expected. Second, for a relatively small $\Delta t_{\rm jet}$, e.g. 0.01s and 0.1s, 
the material beyond the initial jet opening angle ($10^{\rm o}$) is significantly faster than the case
of a large $\Delta t_{\rm jet}$, e.g. 0.5s and 1s. This is because the jet very quickly breaks out from the
ejecta and subsequently forms an angular structure. The line-of-sight material is dominated by the jet material. 
Finally, for a relatively large $\Delta t_{\rm jet}$, e.g. 0.5s and 1s, $dE/d\Omega$ at a large viewing angle 
is large compared with the case of a small $\Delta t_{\rm jet}$, e.g. 0.01s and 0.1s. This, combined with
a relatively small $\bar{\Gamma} \bar{\beta}$, indicates significant mass loading. The line-of-sight material
at a relatively large $\theta_v$ is dominated by the cocoon material. 
In general, the cocoon emission becomes progressively important as $\Delta t_{\rm jet}$ becomes larger,
say, longer than 0.5s.

\section{The case of GW170817/GRB 170817A}

The first NS-NS merger GW event GW170817 was associated with a low-luminosity short GRB 170817A. There are several open questions related to the physics of short GRB 170817A: 1. There was a $\Delta t \sim$1.7s delay of GRB 170817A with respect to GW170817. What is the origin of the delay? 2. The afterglow and prompt emission data are consistent with a structured jet, what is the origin of the jet structure? In particular, is the mildly relativistic material along the line of sight from the cocoon or the wing of a structured jet? 3. What is the radiation mechanism of the $\gamma$-rays, thermal emission from the photosphere or synchrotron radiation? 4. What is the central engine of GRB 170817A, a black hole formed after a brief hypermassive neutron star phase or a long-lived neutron star?

The current available data are not enough to fully address these open questions. Our simulations shed light into some of these problems. According to our simulations, there could be two scenarios to account for the data in principle.

The first scenario, which has been discussed in the literature \citep[e.g.,][]{Nakar18}, interprets the prompt emission as the thermal emission of the cocoon material at shock breakout. This scenario corresponds to the case of a relatively long $\Delta t_{\rm jet}$ (e.g., our M0.5 and M1.0 scenarios). In this model, since the jet launching is delayed, the cocoon emission is significant. Within this scenario, 
the delay time between the jet breakout and emission is
\begin{equation}
\delta t_R \sim R_{\rm GRB} / 2 \Gamma^2 c \approx  0.7~\mathrm{s} \left( \frac{R_{\rm GRB}}{10^{12} \mathrm{cm}} \right) \left( \frac{\Gamma}{5} \right)^{-2},
\end{equation}
which is smaller than the observed 1.7 s delay. One needs to attribute the most delay time to $t_{\rm bo}$, which is consistent with the requirement of a significant $\Delta t_{\rm jet}$. \cite{Meng18} showed that the thermal radiation from a structured jet can account for the observed GRB spectrum. Within this scenario, one has to explain why $\Delta t_{\rm jet}$ is significantly longer than the dynamical time scale of the central engine ($\sim$ millisecond). One possibility is that this time scale is the existence time scale of the hyper-massive NS, and the jet launching happened after the collapse of the NS. This hypermassive NS phase seems to be favored to interpret the kilonova data \citep{Margalit17}. On the other hand, there is no obvious reason why a relativistic jet cannot be launched during the hypermassive NS phase. Another issue of this interpretation is that the observed duration of the short GRB is much longer than $\delta t_{\rm R}$, which defines the typical duration of a shock breakout GRB through the angular spreading time scale. One needs an additional mechanism to interpret the duration.

The second mechanism interprets $\gamma$-ray emission as synchrotron radiation in an optically thin region well beyond the photosphere radius. The thermal emission is suppressed since the jet is Poynting flux dominated \citep{Zhang11}. Within this scenario, $R_{\rm GRB}$ and $\Gamma$ is not specified, but the parameter $\delta t_R \sim R_{\rm GRB} / 2 \Gamma^2 c$ (which depends on both $R_{\rm GRB}$ and $\Gamma$)
is set to a value $\sim 2$ s, which is consistent with both the delay time and the duration of the GRB \citep{ZhangBB18}. Within this scenario, $\Delta t_{\rm jet} \ll \Delta t$, so that the cocoon emission is not significant. The line-of-sight emission is dominated by the wing of the structured jet after the breakout time. \cite{Meng18} showed that synchrotron radiation from a large emission radius \citep{Uhm14,Burgess18} can also interpret the data well. The broadband afterglow emission is also consistent with such a structured jet model \citep[e.g.,][]{Lazzati18,Troja18,Piro19}. Within this scenario, a relativistic jet is launched shortly after the merger, within several dynamical time scales, regardless of the central engine of the short GRB. A black hole may be formed, but not required. A long-lived neutron star can also be the engine of GRB 170817A, as suggested by several authors \citep[e.g.,][]{Yu18,Li18,Ai18,Geng18,Piro19}.

\section{Conclusions}

In this paper, we have performed a series of 2.5D simulations of the propagation of a hot magnetic jet 
in the dynamical ejecta of a binary NS merger.
The effect of the time delay between the merger time and the jet launching time
has been investigated.
Regardless of $\Delta t_{\rm jet}$, a structured jet with an
angle-dependent energy and Lorentz factor is formed after breakout.
The angular distribution of $\bar{\Gamma} \bar{\beta}$ and $dE/d\Omega$ show 
that for a relatively small $\Delta t_{\rm jet}$, the ejecta along the direction of $\theta_v$ greater
than $\theta_j$ is dominated by the jet itself in a low-luminosity, low-$\Gamma$ wing. For a relatively large $\Delta t_{\rm jet}$, e.g. 0.5 s and longer, the large viewing angle direction is dominated by a mildly relativistic cocoon.

Our results suggest that the observed $\sim$ 1.7 s delay between GRB 170817A and the merger time of GW170817
could be explained by either synchrotron scenario with a negligible $\Delta t_{\rm jet}$ or the photosphere scenario
with a relatively large $\Delta t_{\rm jet}$.
Whether the line-of-sight emission is from the wing of a structured jet or the cocoon material depends on $\Delta t_{\rm jet}$.
The data of GW170817/GRB 170817A cannot differentiate between the two scenarios. However, future more GW/GRB associations for NS-NS mergers can help to solve the problem. In particular, the synchrotron scenario requires that the observed delay time scale is comparable to the duration of the burst, while the cocoon scenario interprets the delay time scale and duration with different mechanisms so that the two time scales can be in principle very different. 
Accurate constraints on the jet components and emission radius \citep[e.g.,][]{Matsumoto19} 
provide another way to discriminate these two scenarios. 

In contrast with the simulations of a pure hydrodynamic jet ($\sigma = 0$, e.g., \citealt{Lazzati17,Xie18}) 
or a Poynting-flux dominated jet ($\sigma \ge 1$, e.g., \citealt{Bromberg18}),
we have focused on the hot, magnetic jet with $\sigma \sim 1$ in this work.
A more realistic investigation on the jet structure should include 
both the jet launching mechanism \citep{Kathirgamaraj19},
the energy dissipation within the jet and its propagation in the ejecta.
Furthermore, radiation transfer should be properly implemented in RMHD simulations
to directly relate jet simulations to GRB prompt emission.
All these will be considered in further studies.

\acknowledgments
We thank the anonymous referee for valuable suggestions.
This work is partially supported by the National Natural Science Foundation of China 
(Grants No. 11873030 and 11833003) (JJG and YFH).
JJG acknowledges the supports from the National Postdoctoral Program for Innovative Talents (Grant No. BX201700115), 
and the China Postdoctoral Science Foundation funded project (Grant No. 2017M620199). 
AK and RK acknowledge financial support via the Emmy Noether Research Group on Accretion Flows and Feedback 
in Realistic Models of Massive Star Formation funded by the German Research Foundation (DFG) 
under grant no. KU 2849/3-1 and KU 2849/3-2.
YFH is also supported by the Strategic Priority Research Program of the Chinese Academy of Sciences 
``Multi-waveband Gravitational Wave Universe'' (Grant No. XDB23040400).
The numerical calculations in this paper have been done on the computing facilities 
in the High Performance Computing Center (HPCC) of Nanjing University.

\begin{deluxetable}{cccccccc}
\tabletypesize{\scriptsize}
\tablewidth{0pt}
\tablecaption{Initial Conditions of the Jet Scenarios\label{TABLE:1}}
\tablehead{%
        \colhead{Jet scenario}           &
        \colhead{$\Delta t_{\rm jet}$}   &
        \colhead{$\theta_j$}             &
        \colhead{Luminosity ($L_j$)}     &
        \colhead{$\Gamma_{\infty}$}      &
        \colhead{$v_j^r$}                &
        \colhead{$B^r$}                  &
        \colhead{$B_{j,m}^{\phi}$}       \\ 
        \colhead{}   &   \colhead{(s)}   &  \colhead{}   &  \colhead{(erg s$^{-1}$)}    &  \colhead{}  & \colhead{($c$)} &  \colhead{(G)}  &   \colhead{(G)} 
		}
\startdata
M0.01  & 0.01 & $10\degree$ & $5 \times 10^{51}$ & 265  & 0.8 & $3.1 \times 10^{12}$ &  $6.2 \times 10^{12}$  \\
M0.1   & 0.1 & $10\degree$ & $5 \times 10^{51}$  & 265  & 0.8 & $3.1 \times 10^{12}$ &  $6.2 \times 10^{12}$  \\
M0.5   & 0.5 & $10\degree$ & $5 \times 10^{51}$  & 265  & 0.8 & $3.1 \times 10^{12}$ &  $6.2 \times 10^{12}$  \\
M1.0   & 1.0 & $10\degree$ & $5 \times 10^{51}$  & 265  & 0.8 & $3.1 \times 10^{12}$ &  $6.2 \times 10^{12}$  
\enddata
\end{deluxetable}

\end{document}